\documentclass{article}[12pt,a4paper]
\usepackage[utf8]{inputenc}
\usepackage[english]{babel}
\usepackage[margin=1in]{geometry}
\usepackage{enumitem}
\usepackage[dvipsnames]{xcolor}
\usepackage{amssymb}
\usepackage{amsmath}
\usepackage{amssymb}
\usepackage{bbm}
\usepackage{physics}

\usepackage[
colorlinks=true, 
linkcolor=black, 
citecolor=red, 
filecolor=cyan, 
urlcolor=magenta 
]{hyperref}
\usepackage{braket}
\usepackage{mathrsfs}
\usepackage{eqparbox,xparse,amsmath,amsfonts}
\usepackage{mathtools}

\usepackage[most]{tcolorbox}
\usepackage{float}
\usepackage{graphicx}
\usepackage[shortcuts,acronym]{glossaries}
\usepackage{parskip}
\usepackage[frozencache, cachedir=minted-cache]{minted}
\usepackage{subfigure}
\usepackage[backend=biber, style=ieee]{biblatex}
\usepackage{mathtools}
\usepackage{tikz}
\usepackage{caption}
\usepackage{newfloat}

\newcommand{\py}[1]{\mintinline{Python}{#1} } 

\DeclareFloatingEnvironment[
    fileext=loc,
    listname=List of Codes,
    name=Code,
    placement=H,
]{codeblock}

\newacronym{api}{API}{Aplication Programming Interface}
\newacronym{des}{DES}{Discrete Event Simulation}
\newacronym{gui}{GUI}{graphical user interface}
\newacronym{sim}{QuReed}{}
\newacronym{qunetsim}{QuNetSim}{Quantum Network Simulator}
\newacronym{pip}{pip}{PIP installs packages}
\newacronym{quisp}{QuISP}{Quantum Internet Simulation Package}
\newacronym{omnet++}{OMNeT++}{OMNeT++}
\newacronym{qnic}{QNIC}{...}
\newacronym{sequence}{SeQueNce}{Simulator of QUantum Network Communication}
\newacronym{squanch}{SQUANCH}{Simulator for QUAntum Networks and CHannels}
\newacronym{netsquid}{NetSquid}{The Network Simulator for Quantum Information using Discrete events}
\newacronym{povm}{POVM}{Positive Operator-Valued Measure}
\newacronym{bpsk}{BPSK}{Binary Phase Shift Keying}
\newacronym{netqasm}{NetQASM}{low-level instruction set architecture for quantum internet applications}

\usetikzlibrary{shapes.misc, positioning}
\usetikzlibrary {arrows.meta} 
\newtheorem{definition}{Definition}
\graphicspath{{./assets/beamer/}}

\title{QuReed}
\author{Simon Sekav\v{c}nik, Kareem H. El-Safty, Janis N\"otzel}
\date{May 2024}

\begin{document}

\maketitle

\begin{abstract}
We present \acs{sim}, an open-source quantum simulation framework designed to bridge
gaps between quantum theory, experimental community and engineering. With Quantum
Mechanics maturing and holding significant potential beyond quantum computing,
the need for physically accurate simulations becomes critical. \acs{sim} offers
peer-reviewed simulation models, providing researchers and engineers with
reliable tools for exploring quantum communication protocols and
applications. By facilitating cross-talk between theory and experiments, \acs{sim}
aims to accelerate progress in the field and unlock the transformative power of
quantum mechanics in the communications industry. Its user-friendly Python
interface and comprehensive documentation ensure widespread accessibility and
usability, making \acs{sim} a valuable resource for advancing quantum communication
technologies. 
\end{abstract}


\section{Introduction}

Quantum technologies receive much attention lately, with applications in a variety of domains from computing over sensing to data networks. In particular, its increasing technological level of maturity brings with it a wave of interest in applications utilizing the new possibilities. While previous research was mostly focused on demonstrating novel concepts in principle, the current interest is simultaneously on concrete applications. At this point, system design requires end-to-end thinking, which typically incorporates people of very heterogeneous backgrounds working together to design new devices for the purpose of providing a specific service. In order to efficiently understand the major bottlenecks in such application design, rapid software prototyping based on components that model both theoretically optimal behavior and at the same time the current hardware limitations is a must. In addition, the complexity of the topic may require coordination over different physical locations, where application design needs to be able to formulate specific questions for highly specialized team members. 

While several software tools for quantum system simulation have been developed in the past, none of them has yet targeted the specific demands outlined above. With \acs{sim}, we present a software that is designed for application-specific software prototyping with a scientific performance guarantee.

\subsection{State of the art}
\subsubsection{Quantum Photonics}
In particular the field of quantum photonics has witnessed significant advancements in software packages that facilitate the design, simulation, and implementation of photonic quantum systems. These software tools are crucial for exploring the complexities of photonic quantum computing, quantum communication, and quantum sensing.

One of the leading software packages is \texttt{Strawberry Fields} developed by Xanadu. \texttt{Strawberry Fields} is an open-source library for continuous-variable quantum computation using photonic circuits. It offers a suite of tools for simulating photonic quantum circuits, including support for Gaussian states, non-Gaussian states, and various measurement schemes such as homodyne and heterodyne detection \cite{strawberryfields}. The library integrates seamlessly with machine learning frameworks like TensorFlow, enabling hybrid quantum-classical workflows.

Another prominent package is \texttt{Perceval}, developed by Quandela. \texttt{Perceval} provides a comprehensive platform for simulating and optimizing photonic quantum circuits. It includes features for the manipulation of quantum states, gate operations, and measurement processes. Additionally, \texttt{Perceval} supports the simulation of imperfections in photonic components, which is essential for realistic modeling of experimental conditions \cite{Quandela2023}.

The \texttt{QuTiP} (Quantum Toolbox in Python) is also noteworthy, despite being more general in scope. It has extensive functionalities for simulating open quantum systems, which can be applied to photonics. \texttt{QuTiP} allows for the simulation of master equations and quantum trajectories, providing valuable insights into the decoherence and dissipation processes in photonic systems \cite{Johansson2012}.

\texttt{CQC}'s \texttt{t|ket\textgreater} framework is another significant contribution. Originally designed for superconducting qubits, it has been extended to support photonic qubits and continuous-variable systems. \texttt{t|ket\textgreater} offers efficient circuit optimization routines, which are crucial for minimizing gate operations and enhancing the fidelity of photonic quantum circuits \cite{Sivarajah2020}.

Furthermore, the \texttt{Psi4Numpy} package, although primarily aimed at quantum chemistry, includes functionalities that are useful for simulating the interactions of photons with molecular systems. This cross-disciplinary tool can be leveraged to study photonic processes at the quantum mechanical level \cite{Smith2020}.

In addition to these software tools, several specialized packages focus on niche aspects of photonics. For example, \texttt{MEEP} (MIT Electromagnetic Equation Propagation) is widely used for simulating electromagnetic wave propagation, which is fundamental for designing photonic devices and components \cite{Oskooi2010}. \texttt{Photon-HDF5} is an example that caters to specific needs in photonic quantum information processing and data management, respectively \cite{Photon2022}.

\subsection{Quantum Network Simulators}
In contrast to state of the art quantum network simulators, \acs{sim} is intended to be used in a bottom-up design approach where individual components and their control hardware are the focus. Nonetheless, nontrivial quantum network simulations can be carried out with \acs{sim} as well. There are already a few software packages with the goal of simulating quantum networks.

The researchers constructing classical networks have frequently relied on various classical network simulators. Although the envisioned quantum network is yet to materialize, efforts have been made to simulate quantum processes in a networked manner. In this section, we introduce the key quantum network simulators that address the challenges of quantum networking and computational demands from diverse perspectives. We cover the projects, which are published and the software is available to use either in open-source or closed-source form.

\subsubsection{QuNetSim}
\gls{qunetsim} is an open-source framework built with Python. The Python package is available on the \gls{pip} repositories with very low configuration friction. The repository works out-of the box with the built-in backend. When one needs to make use of any other implemented framework, one needs to also install the backend framework, which is also available on the \gls{pip} repositories.

\gls{qunetsim}\cite{diadamo2021qunetsim} is a flexible, open-source Python tool for quantum network simulations, integrating with backends like Qiskit, ProjectQ, SimulaQron, and EQSN. Its modular, layered approach mimics classical networks, dividing operations into physical, network, transport, and application layers. This enables simulation of quantum networking protocols at different abstraction levels, with qubits managed by backend frameworks and represented as objects.

Available on \gls{pip}, \gls{qunetsim} requires minimal setup and supports hardware-in-the-loop simulation \cite{qns_hong_ou_mandel}. It allows researchers to effectively simulate and analyze quantum networks, taking into account the resource demands of large-scale qubit simulations.

\subsubsection{QuISP}
\gls{quisp}\cite{satoh2022quisp} is an open-source package leveraging the discrete event simulator \gls{omnet++}\cite{varga2010omnet++}, primarily used for network simulations. \gls{quisp} allows researchers to simulate large-scale quantum networks, up to 100 independent networks with 100 nodes each, using a novel approach to quantum state representation called Error Basis. This method tracks deviations from the desired state as an error vector, complementing the available stabilizer states representation, which efficiently describes a subset of quantum states.

QuISP accurately captures quantum systems' behavior with specific quantum state representations and error models, enabling the study of quantum states' impact and errors on network performance. By focusing on deviations from the ideal state, it simulates large networks, excluding general distributed quantum computing simulations due to the non-universality of stabilizer states for all quantum operations. Integrating \gls{omnet++} provides extensive features and robust simulation capabilities, leveraging its vast library of network components and models.

QuISP divides quantum network operations into four layers. The Physical layer includes quantum channels, memories, gates, measurements, and Bell state analyzers. The Link Layer manages connections between adjacent nodes via \gls{qnic} and supports entanglement generation and swapping protocols. The Network Layer encompasses quantum repeaters, routers, routing protocols, and connection setup protocols. Finally, the Application layer provides protocols for end-to-end quality analysis of quantum connections. Overall, QuISP offers a powerful tool for simulating and analyzing quantum communication protocols and their performance under various conditions.

\subsubsection{SeQUeNCe}
\gls{sequence}\cite{wu2021sequence} is an open-source, customizable discrete event framework for simulating quantum networks. Framework is written in Python as well as c++, but end user is meant to use it through Python interfaces. It aims to simulate the networks with the breadth of current and future hardware technologies and protocols. It takes a minimally opinionated approach to the design of the networks, recognizing a lack of consensus for future quantum networks. This means that the current layering of the classical networks is omitted. Nevertheless, the authors introduce a modularized design consisting of five modules: hardware models, entanglement management protocols, resource management, network management, and application module. 

The hardware models module provides the hardware models for the quantum network devices. This module models different devices, with the aim of accurately representing the operation characteristics of the selected simulated hardware, such as: quantum channels, classical channels, quantum gates, photon detectors, and quantum memories. The entanglement management module provides models of entanglement management protocols for the quantum network devices. This module exposes an interface for instantiating and terminating entanglement protocols, which is further used by the resource management module. It also provides interfaces for event notification, notifying of events like photon detection or entanglement expiration. Entanglement management module provides the protocols like \textit{Barret-Kok entanglement generation protocol}\cite{barrett2005efficient}, \textit{BBPSSW} purification protocol\cite{dur2007entanglement} and \textit{entanglement swapping protocol}\cite{zukowski1993event}. The Resource management module manages local resources of \gls{quisp}. It manages local quantum memories according to predetermined rules. Lastly, the Network management module is tasked with the reservation and routing.

\subsubsection{NetSquid}

The focus of \gls{netsquid} lies in quantum circuit construction with a comprehensive set of quantum gates and custom gate definitions, enhanced by C extensions via Cython and the pyDynAA simulation engine. Its noise modeling capability enables realistic emulation of quantum hardware, aiding in error propagation studies and algorithm optimization.

The API supports efficient simulation orchestration, with precise control over parameters, resource allocation, and parallelization, integrating well with high-performance computing environments. As a proprietary tool, NetSquid offers advanced simulation techniques valuable for researchers and developers working on quantum algorithm implementation and performance under realistic conditions.

\subsubsection{SQUANCH}
\gls{squanch}\cite{squanch} is an open-source quantum network simulator written in Python and NumPy. It allows simulation of quantum networks with customizable error models and noisy quantum channels. The framework introduces "agents" as network nodes that can manipulate distributed quantum states and uses computationally optimized methods for handling large streams of quantum information. The entire simulation is fully parallelized, with network nodes running in separate processes, ensuring ease of use and rapid development.

The core class in \gls{squanch} is "QSystem," which manages multi-body maximally entangled quantum systems represented by density matrices implemented as NumPy arrays. Each qubit, represented by the "Qubit" class, references the parent QSystem and has an index number. Qubit measurements yield results in the computational basis, with outcomes determined by the quantum state's probability distribution. This design ensures an intuitive and efficient framework for simulating complex quantum networks.

\subsubsection{SimulaQron}
SimulaQron\cite{dahlberg2018simulaqron} is an open-source software framework for developing and testing quantum communication protocols and algorithms. It provides a simulation environment that mimics real quantum networks, focusing on the classical aspects of networking to create a hardware-independent interface for quantum processing systems. A key feature of SimulaQron is its support for decentralized quantum simulation, enabled by integrating \gls{netqasm}, which manages the localization of quantum simulations on a single machine.

SimulaQron uses various backends to simulate quantum effects, including a built-in stabilizer backend, and supports integration with QuTip and ProjectQ. The framework is designed for easy addition of new backends. However, development activity has slowed, with the latest update in December 2021, and it currently supports Python 3.8, while the latest Python version is 3.11.

\subsubsection{ReQuSim}
ReQuSim\cite{wallnofer2022requsim} is another simulator which falls in a domain of quantum network simulators. ReQuSim is implemented in python and is available on \gls{pip} repository. ReQuSim framework focuses solely on simulation of quantum repeaters, which are poised to represent the basic building block of the future quantum networks. ReQuSim distinguishes itself from other simulators by making use of Monte-Carlo methods as well as \gls{des} operating principles. With these tools ReQuSim attempts to accurately model arbitrary noise process, which can occur in the quantum repeater networks.

\subsection{Comparison of the Simulators}
Each of the presented simulators was optimized for different purpose. Some simulators have chosen to implement the continuous simulation (QuNetSim, SQUANCH, SimulaQron, SeQueNCe) while other have chosen the \gls{des} approach (NetSquid, QuISP, ReQuSim). \gls{des} approach makes processes time aware, where processes are scheduled according to how they would occur in reality, whereas continuous simulation the events happen whenever they are requested, which removes the accurate time description from the simulation.

Further more some simulators chose to heavily specialize for specific use case. QuISP is built on top of existing network simulator, omitting the possibility of simulating arbitrary quantum state. The design choice which pays of as performance increase for the quantum states which can be simulated. Another example of specialization is ReQuSim, which specializes for the simulation of quantum repeater networks, allowing users to precisely analyze different protocols ran on quantum routers.

\
\section{Framework Architecture of \gls{sim}: Devices, Ports and Signals}
In this section we present the \gls{sim}\cite{qureed} simulation framework. \gls{sim} differentiates from other existing simulators, by focusing heavily on the optical state simulation. Already available simulators focus on implementing different functionalities in the confines of the qubit Hilbert spaces, whereas \gls{sim} gives the user the ability to represent optical states. Further more, it introduces the notion of modular, inter-connectable devices which perform quantum (or classical) operations with quantum or classical signals. This design choice, gives the user the ability to simulate complex experimental systems with fine-grained control, while abstracting away tedious management of quantum states.

\begin{minipage}[t]{0.5\textwidth}
\vspace{0pt}
The \gls{sim} framework is designed to facilitate robust interactions between different software components. The section outlines the architectural choices underlying the \gls{sim} framework, which is structured around three fundamental elements: Devices, Ports, and Signals. These elements ensure that the framework supports a modular and extendable approach to system design.

Each component within the framework architecture relies on the concept of the \py{AbstractBaseClass}. This approach mandates that app properties and methods, along with their signatures, must be properly defined and implemented by any extending class. This foundational requirement ensures consistent and predictable behaviour across different components.
\end{minipage}
\hfill
\begin{minipage}[t]{0.45\textwidth}
\vspace{0pt}
    \centering
    \includegraphics[scale=1]{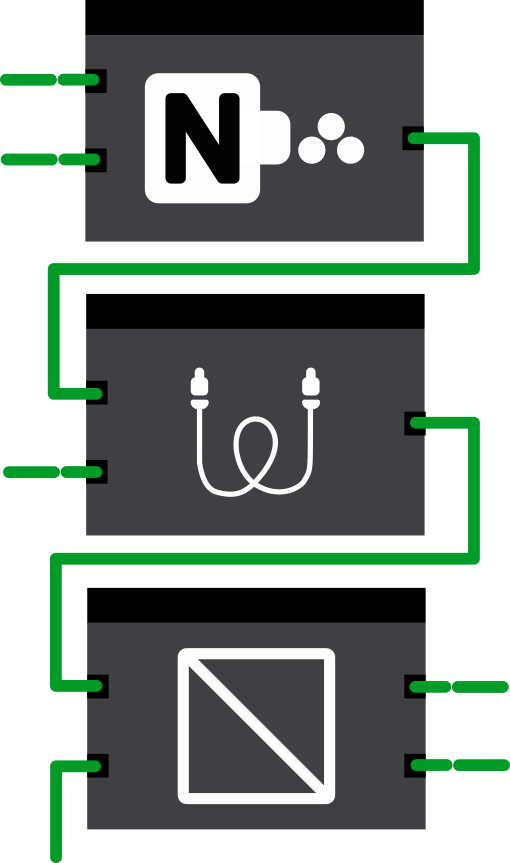}
    \captionof{figure}{Illustration of interconnected Devices, Ports and Signals in the \gls{sim} Framework.}
    \label{fig:enter-label}
\end{minipage}

\subsection{Devices}
Devices in the \gls{sim} framework represent individual units of functionality or modules within a larger system. In principle a device in the \gls{sim} framework serves as a digital twin to a device one might find on an optical table in a laboratory. All implemented devices must extend an existing class \py{GenericDevice}. This design choice enforces the implementation of specific methods and properties. By doing this each device can be characterized by its ability to perform specific functions encapsulated within a well-defined interface. This design allows devices to operate independently and in interconnect with other devices, provided the interfacing allows for their interoperability.

The principle design feature of devices is that they define their \gls{api} through ports. This separation not only isolates the internal implementation of the device from its external interactions but also streamlines system integration and scalability.

\subsection{Ports}
Ports are the conduits through which devices expose their functionalities and communicate with external environment or other devices. Each port in the \gls{sim} framework is designed to serve as a gateway that strictly regulates the type of data or more precisely \py{Signal} it can transmit or receive. This rigorous specification ensures that the interactions between devices adhere to predefined protocols, enhancing the reliability and predictability of the entire system.

The design of ports emphasizes compatibility and interoperability, allowing devices from different domains or with varying functionalities to connect and function cohesively. This is achieved by standardizing the signal types that can be accepted by a port, thereby enforcing consistency across the framework.

\subsection{Signals}
Signals in the \gls{sim} framework are the predefined units of information transmitted through ports between devices. These signals are critical for maintaining the integrity and security of data exchanges within the system. The framework ensures that all communications adhere to a consistent format and semantic, facilitating error-free data transmission and processing.

To accommodate the specific needs of various applications, the \gls{sim} framework allows for the implementation of custom signals. This flexibility is crucial for adapting the system to diverse operational requirements and enhancing its applicability across different technological and research domains.Implementing custom signals involves defining new data types and ensuring that these types are compatible with the existing ports and devices in the framework.

Custom signals must adhere to the framework's strict guidelines on data formats and property definitions. The adherence is essential to prevent conflicts and errors in data handling, ensuring that even custom implementations integrate seamlessly with the core functionality of the \gls{sim} system. By allowing custom signals, the framework not only enhances its flexibility but also its capability to handle specialized tasks that are not covered by the standard signal types.

The definition and management of signals, whether standard or custom, are tightly controlled to optimize performance and minimize overhead. This control makes the framework efficient and effective in handling complex interaction between diverse devices.

\subsection{Signal Inheritance and Port Compatibility}
In the \gls{sim} framework, signals are designed with an inheritance structure that facilitates flexibility and extensibility. This design allows for the creation of specific signal types that extend from more generic ones, adhering to object-oriented principles that promote reusability and scalability.

For instance, consider a hierarchy where \py{GenericSignal} serves as the base class for all signals. A subclass \py{GenericQuantumSignal} extends \py{GenericSignal}, specializing it for quantum-based applications. Further specializing, \py{PhotonicQuantumSignal} might extend \py{GenericQuantumSignal}, indicating a more specific type of quantum signal that deals with photonic processes.

Under this structure, a port designed to accept a \py{GenericQuantumSignal} can inherently accept any of its subclasses, such as \py{PhotonicQuantumSignal}. This compatibility is due to the principle of polymorphism in object-oriented programming, where a port that accepts a particular class of signal can also accept any of its derivatives. This approach not only simplifies the system design but also enhances its adaptability, allowing new types of signals to be integrated seamlessly into the existing framework.

This model ensures that the system remains robust and versatile, capable of evolving with technological advancements and varying scientific needs without requiring substantial modifications to the core architecture. Ports that accept generic signals can handle wide array of specific signal instances, promoting greater interoperability across different devices and applications within the framework.

\subsection{Discrete Event Simulation}

The \gls{sim} framework incorporates a \gls{des} engine that provides advanced simulation capabilities for backend systems that utilize the \gls{des} methodology. While the framework supports multiple backends, not all are required to use the \gls{des} approach. However, for those that do, the \gls{des} engine offers a powerful tool for accurately simulating complex interactions between devices within the system with time awareness.

\subsubsection{Operation of the \gls{des} Engine}

The \gls{des} approach is employed in the following manner within the \gls{sim} framework:
\begin{enumerate}
    \item \textbf{Start of Simulation}: When a simulation is ran, a user defines how long in terms of simulated time, the simulation should last. Due to precision needed, the time is denoted in a higher precision, than allowed by built in float type. To track time we make use of \emph{mpmath} library\cite{mpmath}. The simulation concludes, when wither timestamp of next event is larger then set simulation time or the event queue is empty.
    \item \textbf{Initial Event Triggering}: At the start of the simulation devices, which implement a method \py{des_init} are executed. These devices are crucial for setting the initial conditions and starting point of a simulation.
    \item \textbf{Event Handling by Devices}: In order for a device to provide functionality in the scope of \gls{des}, it must implement a method \py{des}. This method is executed by the simulation manager according to the scheduled events and is responsible for producing timestamped outputs. Specifically, the method generates signals and labels these output on their respective outgoing ports accompanied with timestamps indicating when the event at next corresponding device should be executed.
    \item \textbf{Event Scheduling}: The methods of \py{des} and \py{des_init}, must be decorated using provided decorator \py{@schedule_next_event}, which is responsible for finding the next device and scheduling the event, using the timestamp and signal provided by the output of these two functions. In effect, each event is timed according to the specified timestamps ensuring that signals 'reach' destinations precisely when intended.
    \item \textbf{Event Merging}: If a device receives multiple inputs at the same timestamp, these events are automatically merged by the simulation engine. This merging is critical as it allows the device to process multiple simultaneous inputs effectively. 
\end{enumerate}

\subsection{Backends}

The \gls{sim} framework employs a structured approach to implementing and integrating devices, relying on backends to simulate quantum mechanics effectively.
\gls{sim} is equipped with a built-in backend — the \textbf{Fock Backend} — and provides the flexibility to incorporate additional backends, such as Hamiltonian or Gaussian, in future updates. Additionally, \gls{sim} relies heavily on an external backend, \textbf{Photon Weave}, to extend its simulation capabilities.

\subsubsection{Built-in Fock Backend}
The \textbf{Fock Backend} is integral to \gls{sim}, specializing in simulations involving quantum states in Fock space. This backend is tailored to scenarios that require precise manipulation of number states. Unlike other backends, built-in Fock backend does not sequence operations over time. Instead, it prepares the initial quantum states and executes all gates in the right order at once, providing users with post-experiment statistics. This method ensures efficient handling of static quantum simulations where temporal dynamics are not the focus.

\subsubsection{External Photon Weave Backend}
to augment its native capabilities, \gls{sim} integrates with \textbf{Photon Weave} as an external backend. This system encodes temporal modes in both Fock and Polarization spaces. Photon Weave is specifically designed to align with the \gls{des} paradigm, enabling dynamic quantum simulations. This backend proves particularly effective for simulating intricate quantum effects such as Hong-Ou-Mandel effect \cite{bouchard2020two}.

\section{Use Cases}
The use cases covered in \gls{sim} have been systematically aggregated (following the principle of lean software development) based on a list of so-called user stories which can in full detail be found in the appendix.

In this section we bring to light some of the experiments simulated using the \gls{sim}.

\subsection{Information Theoretical Joint Detection}

For first interesting experiment to showcase the capabilities of \gls{sim} we choose information theoretical joint detection data transmission, following the sequential decoding method described in \cite{giovannetti2012} but with an encoding where \gls{bpsk} is used for the generation of signals. The joint detection data transmission displays a logarithmic advantage over conventional data transmission \cite{guha2011quantum}. The transmission  of a message is accomplished by phase modulating a number $P$ of coherent pulses. After receiving the pulses, the receiver tries to decode the message via the sequential decoding method. It iteratively applies a set of phase shifts $(\phi_1, \ldots, \phi_P)$ where each phase shift can take value in $\phi\in\{0,\pi \}$. More precisely, the receiver goes through the list $\{((-1)^{b_0}\pi,\ldots,(-1)^{b_P}\pi)\}_{b1,\ldots,b_P\in\{0,1\}}$ in lexicographic order. In each round, it first applies the corresponding series $((-1)^{b_0}\pi,\ldots,(-1)^{b_P}\pi)$ of phase shifts, followed by execution of a \gls{povm} with two outcomes $Y$ and $N$. If the outcome is $Y$, the receiver declares the message to be the $m$ with binary representation $b_1\ldots b_P$ and stops. If the outcome is $N$, the receiver applies the inverse phase shift $(-(-1)^{b_0}\pi,\ldots,-(-1)^{b_P}\pi)$ and proceeds to the next sequence $((-1)^{b_0'}\pi,\ldots,(-1)^{b_P'}\pi)$ in the list. 

The \gls{povm} measurement is defined with the following operators:
\begin{align}\label{eq:jdr_povm}
M_Y &= \ket{0} \bra{0}^{\otimes P} &
M_N &= \mathbb{I} - M_1.
\end{align}

\begin{figure}[H]
    \centering
    \includegraphics[width=0.8\textwidth]{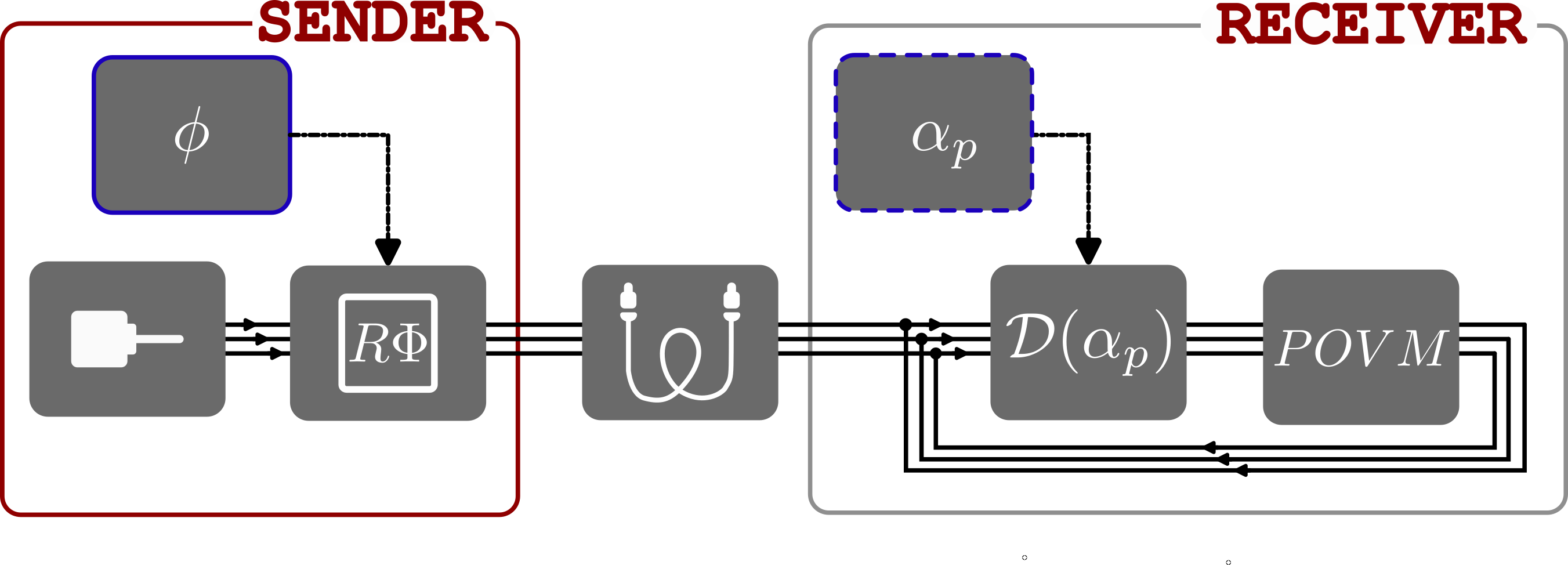}
    \captionof{figure}{Illustrated configuration for the simulation of joint detection receiver. Three pulses are denoted by three separate lines, flowing from the coherent light source. Sender encodes code word by selecting specific phase for each pulse. Receiver displaces each pulse by specific $\alpha_p$, which defines the guess of the code word. Then the pulses enter the \gls{povm} measurement device. In case vacuum was measured, the device has effectively guessed the code word. If vacuum was not measured, then the pulses is fed back into to the displace for another guess.}
    \label{fig:jdr_setup}
\end{figure}

The full data transmission procedure is implemented using \ac{sim}, and the simulation setup is illustrated in Figure \ref{fig:jdr_setup}. We limit our simulation to $P=3$ pulses. The consecutive coherent pulses are created by displacing vacuum state:
\begin{equation}
    \ket{\alpha} =\mathcal{D}(\alpha) \ket{0} = e^{\alpha \hat a^\dagger - \alpha^* \hat a} \ket{0},
\end{equation}
where a value $\alpha=0.4$ is chosen in order to have a low dimensional representation of the system and thereby limit the computational complexity of the simulation. After pulse creation we encode the message $m$ into the pulse by first creating its binary representation $b_1,\ldots,b_P$ and then individually phase shifting the pulses by $\phi_p =(-1)^{b_p}\pi$. 

\begin{figure}[H]
    \centering
    \includegraphics[scale=1]{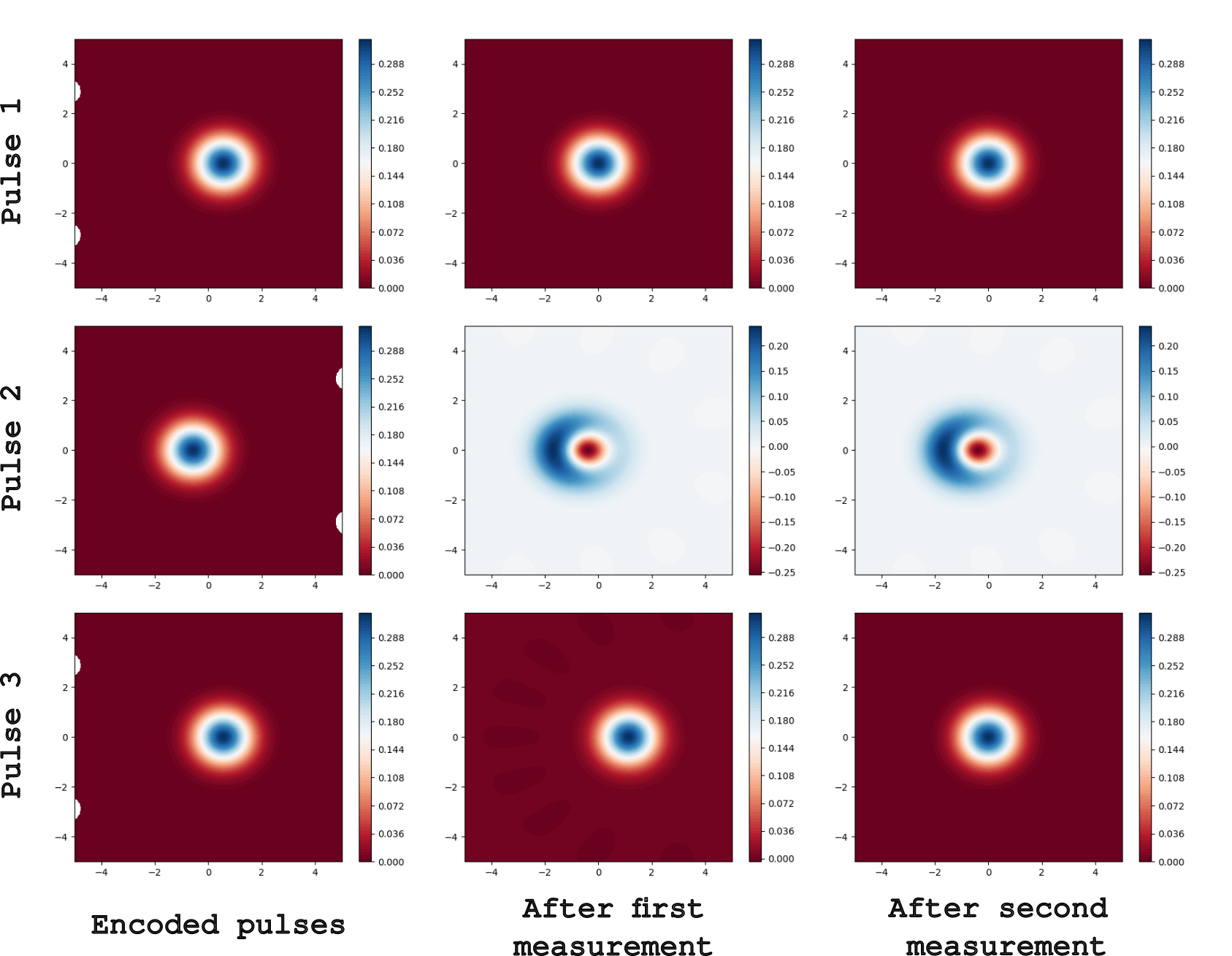}
    \captionof{figure}{Wigner plots of individual three pulses after encoding, after first measurement and after second measurement. On the third measurement, the detector accurately measures vacuum, signaling the correct guess of code word. Due to relatively small $\alpha=0.4$ the pulses are subject to noise as can be seen in the second pulse evolution.}
    \label{fig:jdr_wigner}
\end{figure}

In the Figure \ref{fig:jdr_wigner}, we display the progressive change of the Wigner plots for the three pulses as the decoder tries to guess the message. For this experiment we arbitrarily chose the code word $m=3$. With this code word the measurement device should yield measurement result $Y$ on the third try.

\subsection{Graphical User Interface}

\begin{minipage}[t]{0.47\textwidth}
\vspace{0pt}
The \gls{sim} framework ships with a \gls{gui}, which is a preferred way of interacting with the simulations. The \gls{gui} simplifies management complex simulation setups, by providing an interactive way of configuring the individual components of the simulated experiment.

The \gls{gui} is a critical component that bridges the gap between advanced quantum simulations and user accessibility, ensuring both novice and expert users can effectively utilize the framework. Further more the \gls{gui} enhances collaboration. Instead of collaborating on a lengthy code block, users could immediately understand how devices are interconnected and what are the simulation conditions.

The interface is structured in a way to seamlessly guide users through the workflow of configuring and running each simulation. It gives user intuitive way of implementing custom devices and signals and makes them immediately usable inside of the larger experiment. Implemented custom devices can be dragged from the left panel into any simulation scenario.
\end{minipage}
\hfill
\begin{minipage}[t]{0.47\textwidth}
\vspace{0pt}
\begin{figure}[H]
    \centering
    \includegraphics[width=0.9\textwidth]{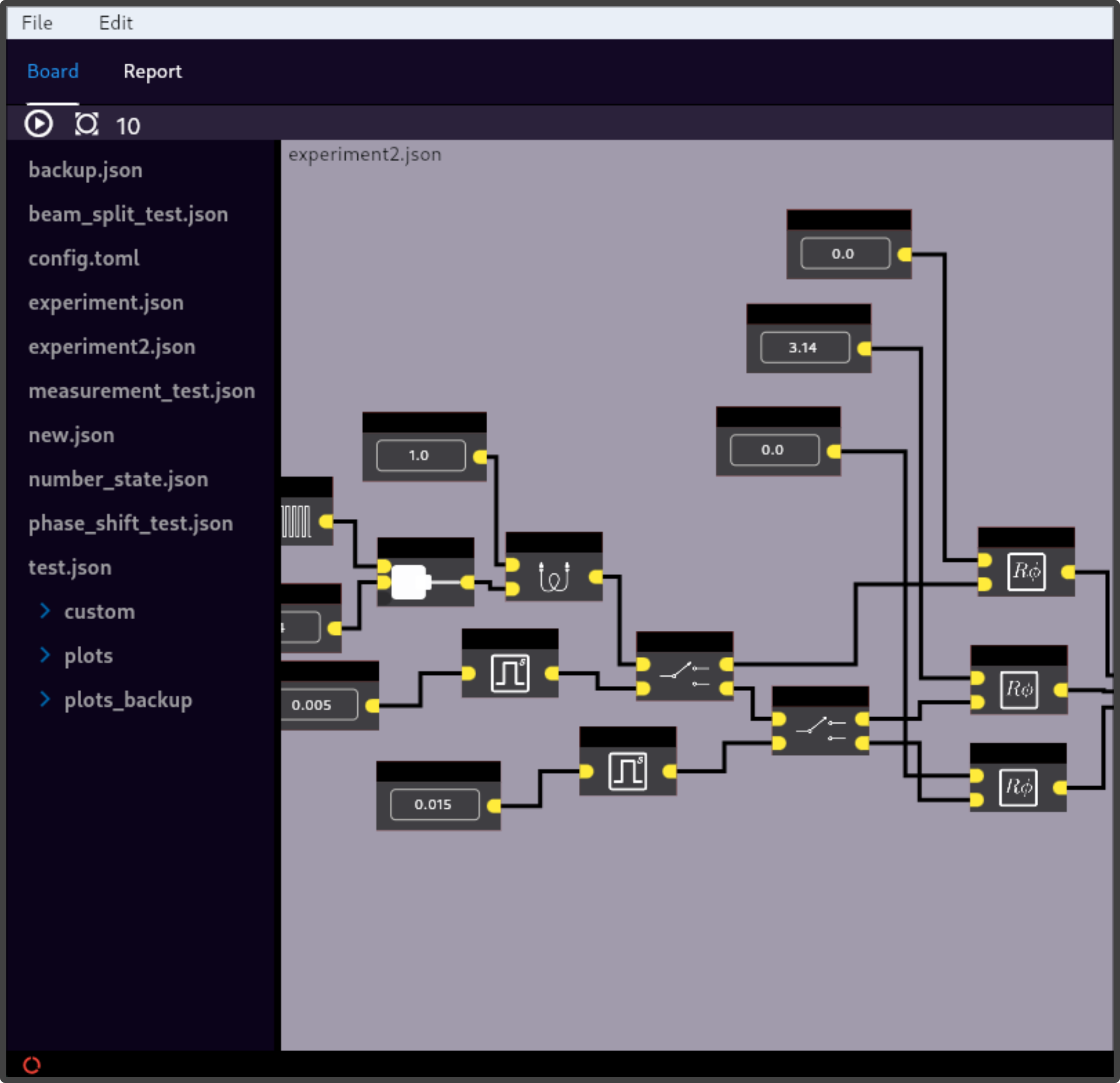}
    \caption{Graphical User Interface of \gls{sim} framework, allows for connecting devices and implementing new custom devices.}
    \label{fig:gui}
\end{figure}
\end{minipage}
\section{Conclusions}

In this work we presented \gls{sim}, which is a modular, robust and user-friendly simulation tool. \gls{sim} bridges the gap between theoretical quantum mechanics and practical engineering applications. Key contributions of this work are:
\begin{enumerate}
    \item Enhanced Simulation Capabilities: \gls{sim} integration of \gls{des} engine allows for precise timing and handling of events, making it possible to accurately model complex interactions between quantum (and classical) devices. This capability is crucial for developing and testing optical quantum communication protocols.
    \item Modular Design: The framework's use of Devices, Ports and Signals ensures a modular approach to system design. This modularity facilitates the creation of complex experimental setups and the easy integration of new components, promoting flexibility and scalability.
    \item Focus on Optical State Simulation: By concentrating on optical state simulation, although not omitting general Hilbert spaces, \gls{sim} addresses a niche that other quantum network simulators often overlook. This design choice is essential for advancing quantum communication technologies.
    \item Comprehensive User Interface: The inclusion of a \gls{gui} enhances accessibility and usability. The \gls{gui} simplifies the process of setting simulation parameters and connecting the devices in desired fashion.
\end{enumerate}

In conclusion, \gls{sim} stands out as a comprehensive and versatile quantum simulation tool that meets the demands of both theoretical exploration and practical application engineering. Its robust architecture, extensive simulation capabilities, and user-friendly interface make it an asset for advancing the field of quantum communication. By facilitating cross-disciplinary collaboration and providing reliable simulation models, \gls{sim} aims to accelerate progress in the field.
\newpage
\section*{Acknowledgement}
This work was financed by the Federal Ministry of Education and Research of Germany via grants 16KIS1598K, 16KISQ039, 16KISQ077 and 16KISQ168 as well as in the programme of ``Souver\"an. Digital. Vernetzt.''. Joint project 6G-life, project identification number: 16KISK002. We acknowledge further funding by the DFG via grant NO 1129/2-1, and by the Bavarian Ministry for Economic Affairs (StMWi) via the project 6GQT and by the Munich Quantum Valley. We acknowledge insightful discussions with our colleagues Paul Kohl, Davide Li Calsi and Shahram Dehdashti, and with Matheus Ribeiro-Sena.
\appendix
\section{Continuous Variable Description}
\subsubsection{Fock Backend}
In quantum mechanics (QM), particles can be categorized as \textit{fermionic} or \textit{bosonic}. A big portion of the literature about QC and the industry focuses on the formalism of the two-level system denoted by the $\ket{0}$ and $\ket{1}$ states and can be efficiently described by a complex Hilbert space denoted by $\mathcal{H}$ spanning all the possible combinations of these states. It's also called the discrete model of QC. The physical realization for such a system would be Superconducting qubits or Ion-Trapped qubits, however other hardware implementations are also considered like $\cdots$. With an appropriate set of operators or gates and enough number of qubits, this formalism can simulate any classical function.\\

\begin{figure}[!ht]
	\centering
	\includegraphics[scale=0.8]{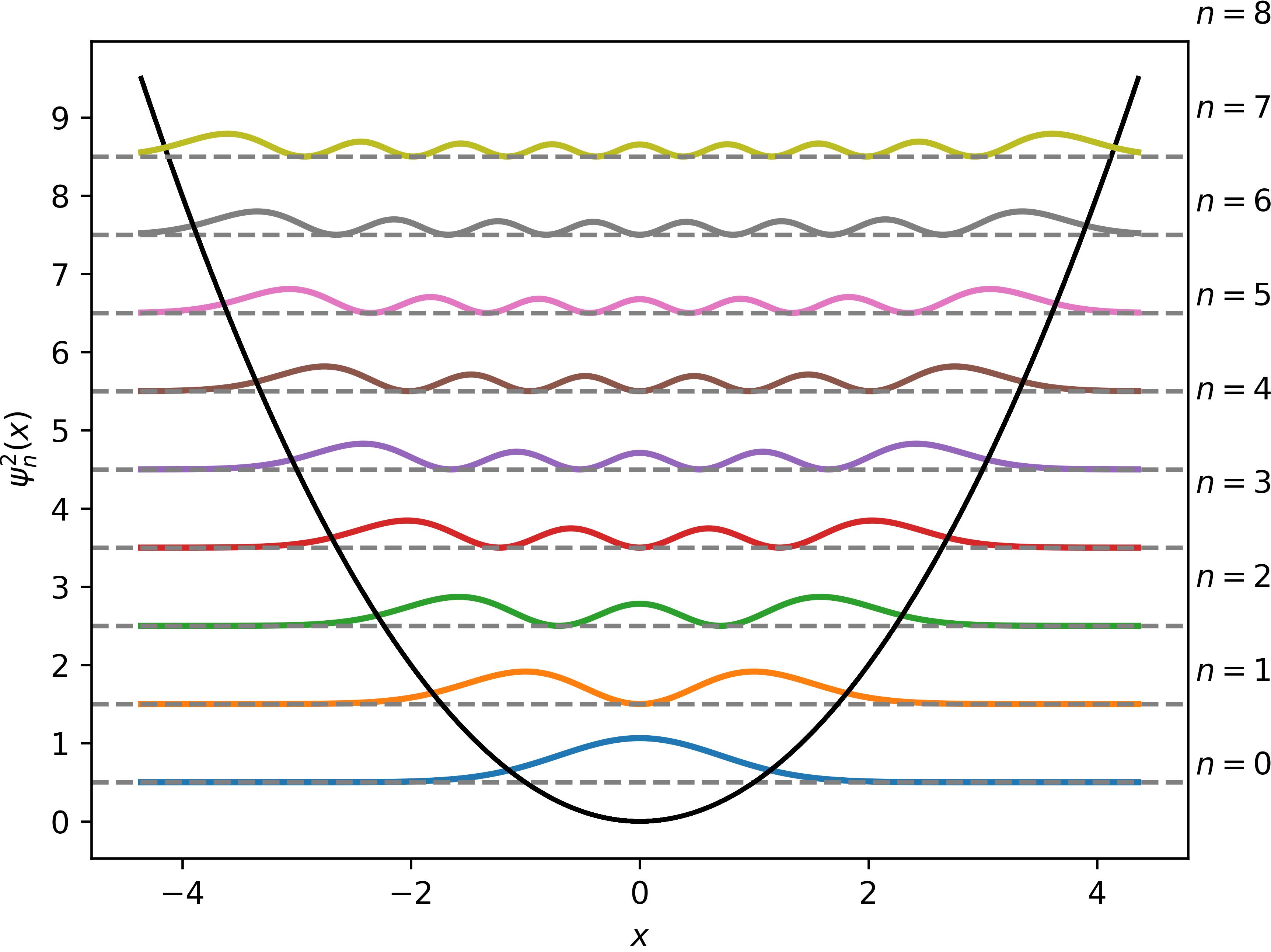}
	\caption{Different energy levels of a Quantum HO. By setting $\hbar$ and $\omega$ to be $1$, equation \eqref{eigenvalues of HO} will be $(n+0.5)$. This is the reason for the shifted energy level at $n=0$. It's also obvious that each energy level is equally spaced.}
	\label{graphic:QHO}
\end{figure}

On the other hand, quantum computers that use bosonic particles such as \textit{photons} to simulate a quantum process or to perform a logical operation, do have a higher degree of freedom in terms of the number of particles that can be within a single state and the fact that we can have a system that is composed of different states and each state has different $N$ particles.
To explain this visually, let's consider a Harmonic Oscillator (HO) with mass $m$ moving inside a quadratic potential energy function of displacement
\begin{equation} \label{v(x)}
    V(x)=\frac{1}{2}kx^2,
\end{equation}
its Schrödinger equation will be
\begin{equation}\label{QHO}
    E \psi(x) = -\frac{\hbar^{2}}{2m} \frac{d^{2}}{d x^{2}}\psi(x) + \frac{1}{2} kx^2 \psi(x).
\end{equation}
The eigenstates of equation \eqref{QHO} can be written in terms of the Hermite polynomials as
\begin{equation}\label{eigenstate}
    \psi_n(x) = \mathcal{N}_n  H_n(x) e^{-x^2/2},
\end{equation}
where $\mathcal{N}_n = (\sqrt{\pi} 2^n n!)^{-1/2}$ is a normalization factor, $H_n$ is the Hermite polynomial of order $n$, and $n \in \mathbb{N}$ is the quantum number. The eigenvalues can be computed as a function of $n$
\begin{equation}\label{eigenvalues of HO}
    E_n= \hbar \omega \Big (n+\frac{1}{2}\Big),
    \end{equation}
where $\omega$ is the angular frequency that is dependent on $k$ and $m$. Figure \ref{graphic:QHO}, depicts the different energy levels and the associated quanta particles in each level for a single quantum HO. In this manner we can describe a single quantum HO as an equivalent to a single qubit. From such analogy, it's obvious that there are much more degrees of freedom in the photonic system than the qubit one has. If there are multiple disjoint quantum HOs for a system, then a special Hilbert space must be introduced to efficiently describe the quantum system.

Named after Vladimir A. Fock, the Fock space was introduced in 1932. For a Hilbert space $\mathcal{H}$ that contains a single particle or a quantum state that could be either $\ket{0}$ or $\ket{1}$, the tensor product $\mathcal{H}\otimes\mathcal{H}$ will be the space that fully characterizes the quantum states of the same two particles within it. One can generalize this to $\mathcal{H}^{\otimes N}=\otimes_{k=1}%
^{N}\mathcal{H}_{k}$ for $n$ particles. Since we focus on photonic QC, we shall focus on the bosonic Fock space which is also called the symmetrized tensor product of Hilbert spaces.

A single mode state can be decomposed into the Fock basis as follows:
\[
\ket{\psi} = \sum_n c_n \ket{n},
\]
if there exists a unique integer $m$ such that
\[
\begin{cases}c_n=1& n=m\\c_n=0&n\neq
m\end{cases},
\]
then the single mode is simply a Fock state or  $n$ photon state.
When working with an $N$-mode density matrix in the Fock basis,
\[
\rho = \sum_{n_1}\cdots\sum_{n_N} c_{n_1,\cdots,n_N}\ket{n_1,\cdots,n_N}\bra{n_1,\cdots,n_N}.
\]
We use the convention that every pair of consecutive dimensions corresponds to a subsystem; i.e.,
\[
\rho_{\underbrace{ij}_{\text{mode}~0}~\underbrace{kl}_{\text{mode}~1}~\underbrace{mn}_{\text{mode}~2}}
\]

\begin{definition}[Bosonic Fock Space]\label{define_1}
    A Bosonic Fock space is the sum of all symmetrized tensor products of a single particle Hilbert space $\mathcal{H}$
    \begin{equation}\label{fock space}
        \mathcal{F} = \bigoplus_{n=0}^{\infty} \mathcal{H}^{\odot n},
    \end{equation}
    and the  symmetrized tensor product $\odot$ is given by $\frac{1}{n!} \sum_{e \in V_n} y_{e(1)} \odot y_{e(2)} \odot \cdots \odot y_{e(n)}$ where $V_n = \{1, 2, \cdots, n\}$ is the set of permutations and $y \in \mathcal{H}$.
\end{definition}
According to definition (\ref{define_1}), if we have a quantum system of $3$ particles and $4$ modes then the state $\psi$ can be written as $|0, 1, 1, 1\rangle$, $|0, 3, 0, 0\rangle$, $|1, 0, 2, 0\rangle$, or any other combination as long as the total number of particles is equal to $3$. It's clear that the Fock space enables us to express the possibility of evaluating the expectation values of finding $N$ particles in a specific eigenstate, i.e., it becomes natural to calculate the mean photon number of a specific quantum state.\\

\subsection{Field Operators}
While developing the second quantization formalism of quantum mechanics, Paul Dirac has introduced the field operators to be used instead of wavefuncitons. There are two important field operators; the annihilation operator and its adjoint the creation operator denoted by $\hat{a}$ and $\hat{a}^\dagger$ respectively. The creation operator or the raising operator is used to increse the number of particles in a quantum system. On the other hand, the lowering operator or annihilation operator is used to remove a particle from a quantum system. These operators satisfy the bosonic commutation relations
\begin{equation}
    \lbrack\hat{a}_{i},\hat{a}_{j}^\dagger] = \hat{a}_{i} \hat{a}_{j}^\dagger - \hat{a}_{j}^\dagger \hat{a}_{i} = \delta_{ij}. \label{BCR}%
\end{equation}
The matrix representation of the creation operator is
\begin{equation}\label{creation operator matrix}
    \hat{a}^\dagger  = \begin{bmatrix}
    0 &  0 & 0 & 0 & \cdots\\ 
    \sqrt1 & 0 & 0 & 0 & \cdots\\ 
    0 & \sqrt2 & 0 & 0 & \cdots\\ 
    0 & 0 & \sqrt3 & 0 & \cdots\\
    \vdots & 0 & 0 & \ddots &  0
    \end{bmatrix}.
\end{equation}
There are also two important operators that represent the position and the momentum of a quantum system denoted by $\hat{x}$ and $\hat{p}$ respectively. They are given as
\begin{equation}\label{position and momentum operators}
    \hat{x} = \sqrt{\frac{\hbar}{2}}(\hat{a}+\hat{a}^\dagger),~~~ \hat{p} = -i \sqrt{\frac{\hbar}{2}}(\hat{a}-\hat{a}^\dagger),
\end{equation}
and they fulfil the commutation relation
\begin{equation}\label{Position and Momentum commutation rule}
    [\hat{x}_i, \hat{p}_j] = i \hbar\delta_{ij}.
\end{equation}

In a Fock space of $0$ particles or a Vacuum state denoted by $\ket{0}$, the effect of the ladder operators $\hat{a}$ and $\hat{a}^\dagger$ is
\begin{equation}\label{equation4}
    \hat{a}\left\vert 0\right\rangle =0,~~~\hat{a}\left\vert n\right\rangle
    =\sqrt{n}\left\vert n-1\right\rangle ~~\text{(for}~n\geq1\text{),}
\end{equation}
and
\begin{equation}\label{equation5}
\hat{a}^{\dagger}\left\vert n\right\rangle =\sqrt{n+1}\left\vert
n+1\right\rangle ~~\text{(for}~n\geq0\text{).}
\end{equation}
Using equation \eqref{equation5} one can create any number state or a Fock state from the Vacuum state as follows
\begin{equation}\label{equation6}
|n\rangle = \frac{(\hat{a}^\dagger)^n}{\sqrt{n!}} |0\rangle.
\end{equation}
The ladder operators can also form a hermitian operator called the \textit{Number Operator} denoted by 
\begin{equation}\label{number operator}
    \hat{n} = \hat{a}^\dagger\hat{a}.
\end{equation}

\subsection{Linear Passive Operators in the Fock Space}

There is a unitary operator called the Beamsplitter. The Beamsplitter operates on two quantum modes instead of one. It's defined as

\begin{equation}\label{beamsplitter}
    \mathcal{B}(\theta,\phi) = \exp\left(\theta (e^{i \phi}\hat{a_1} \hat{a_2}^\dagger - e^{-i \phi} \hat{a_1}^\dagger \hat{a_2}) \right).
\end{equation}
The effect on the ladder operators is given by
\begin{equation}
    \begin{split}
        \mathcal{B}^\dagger(\theta,\phi) \hat{a_1}  \mathcal{B}(\theta,\phi) &= t \hat{a_1} -r^* \hat{a_2},\\ \mathcal{B}^\dagger(\theta,\phi) \hat{a_2}  \mathcal{B}(\theta,\phi) &= t \hat{a_2} +r \hat{a_1},
    \end{split}
\end{equation}
where $t=\cos{\theta}$ is the transmittivity and here $r$ is the reflectivity amplitude and equals $e^{i\phi} \sin{\theta}$. When $\theta = \pi/4$ and $\phi = 0$ or $\phi = \pi$, the resulting operator is a 50-50 Beamsplitter. It's clear that the Beamsplitter is not an active operator since it doesn't increase the photon number in the quantum system.\\

The Rotation operator denoted by $\mathcal{R}(\phi)$ is a single mode operator that has the following transformation on the lowering operator
\begin{equation}
    \mathcal{R}(\phi) = \exp\left(i \phi \hat{a}^\dagger \hat{a}\right).
\end{equation}
It's a passive linear operator that rotates the quadrature operators depending on $\phi$.

\subsection{Coherent States}\label{cs}
    
The Glauber States or as widely known as the Coherent States (CS) of light are specific quantum states of the quantum harmonic oscillator. They can be used to describe the dynamics of a state whose ground state has been displaced from the origin. One of the main applications of CS is laser physics. CS can be generated by applying the Displacement operator \cite{book}
    \begin{equation}\label{displacement operator}
        \mathcal{D}(\alpha) = \exp(\alpha \hat{a}^\dagger - \alpha^* \hat{a})
    \end{equation}
    on a Vacuum state where $\alpha \in \mathbb{C}$ is the value of the complex displacement and $\alpha = r e^{i\theta}$ where $r \in \mathbb{R}$ is the amplitude of the displacement and $\theta \in [0, 2\pi)$ is the phase of the displacement. Setting $\alpha$ to be $0$ will be the same as the Vacuum state $|0\rangle$. In terms of the Fock space representation, the CS can be denoted as
    \begin{equation} \label{CS}
        \begin{split}
            |\alpha \rangle & = \exp(-\frac{|\alpha|^2}{2}) \exp(\alpha \hat{a}^\dagger) \exp(-\alpha^* \hat{a}) |0\rangle \\
            & = \exp(-\frac{|\alpha|^2}{2}) \exp(\alpha \hat{a}^\dagger) |0\rangle  \\
            & = \exp(-\frac{|\alpha|^2}{2}) \sum_{n=0}^{\infty} \frac{\alpha^n}{\sqrt{n!}} |0\rangle.
        \end{split}
    \end{equation}\\

The CS $|\alpha\rangle$ is a right eigenstate of the lowering operator $\hat{a}$ with an eigenvalue $\alpha$ which implies $\hat{a}|\alpha\rangle = \alpha|\alpha\rangle$. On the other hand the CS $|\alpha\rangle$ is a left eigenstate of the raising operator $\hat{a}^\dagger$ which can be denoted by $\langle\alpha|\hat{a}^\dagger = \alpha^*\langle\alpha|$. The former two results can be summarized as
    \begin{equation}\label{cs_eigen}
        \mathcal{D}^\dagger(\alpha) \hat{a} \mathcal{D}(\alpha) = \hat{a} + \alpha \boldsymbol{I},
    \end{equation}
    \begin{equation}\label{cs_eigen_left}
        \mathcal{D}^\dagger(\alpha) \hat{a}^\dagger \mathcal{D}(\alpha) = \hat{a}^\dagger + \alpha^* \boldsymbol{I}.
    \end{equation}
The same can be shown for the quadrature operators $\hat{x}$ and $\hat{p}$
    \begin{equation}
        \begin{split}
        \mathcal{D}^\dagger(\alpha) \hat{x} \mathcal{D}(\alpha) = \hat{x} +\sqrt{2 \hbar } \operatorname{Re}(\alpha) \boldsymbol{I},\\
        \mathcal{D}^\dagger(\alpha) \hat{p} \mathcal{D}(\alpha) = \hat{p} +\sqrt{2 \hbar } \operatorname{Im}(\alpha) \boldsymbol{I}.
        \end{split}
    \end{equation}
Using equations \eqref{cs_eigen} and \eqref{cs_eigen_left}, the expectation value of the number operator $\hat{n}$ can be found as
    \begin{equation}\label{mean photon number coherent}
        \begin{split}
            \bra{\alpha}\hat{n}\ket{\alpha} &= \bra{0}\mathcal{D}^\dagger(\alpha)\hat{n}\mathcal{D}(\alpha)\ket{0}\\
            &=\bra{0} (\hat{a}^\dagger + \alpha^*)(\hat{a} + \alpha)\ket{0}\\
            &=|\alpha|^2
        \end{split}
    \end{equation}
    The inner product $\braket{.,.}$ between two CSs $\ket{\alpha}$ and $\ket{\beta}$ is given as
    \begin{equation}
        \braket{\alpha,\beta} = \exp(-\frac{|\alpha|^2}{2})\exp(-\frac{|\beta|^2}{2}) \sum_{n=0}^{\infty} \sum_{m=0}^{\infty} \frac{\beta^n}{\sqrt{n!}} \frac{(\alpha^*)^m}{\sqrt{m!}},
    \end{equation}
    which can be rewritten in a compact form as
    \begin{equation} \label{cs_inner}
        |\braket{\alpha,\beta}|^2 = \exp(-|\alpha - \beta|^2).
    \end{equation}
    It's clear that equation (\ref{cs_inner}) will evaluate to $1$ if and only if $\alpha = \beta$. In case of having a large difference between the displacement values, equation (\ref{cs_inner}) can be approximated to $0$.
    
\subsection{Squeezed States}
    Discovered by Earle Hesse Kennard, Squeezed light states are a type of non-classical states of light that allow to have low uncertainty or noise below the standard quantum limit in one of the quadrature operators while the other quadrature has an increased uncertainty. Thus, a squeezing operation obeys the Heisenberg principle and introduces a new prospective to the typical CS characteristics that can be considered as a classical Gaussian wave-packet. It's mostly used in Quantum Sensing, Radiometry, and Entanglement-based quantum key distribution. The squeezing operator $S(z)$ is \cite{book}
    \begin{equation}\label{squeezing operator}
         \mathcal{S}(z) =\exp\left(\frac{1}{2}\left(z^* \hat{a}^2-z \hat{a}^{\dagger^2}\right) \right),
    \end{equation}
    where $z = r e^{i\phi}$, with $r \in [0, \infty)$ and $\phi \in [0, 2\pi)$.
    A Squeezed state is generated by applying the squeezing operator to a Vacuum state
    \begin{equation}\label{squeezed state representation}
        \begin{split}
            \ket{z} &= \mathcal{S}(z)\ket{0}\\
            & = \sqrt{\sech r} \sum_{n=0}^{\infty} \frac{\sqrt{(2n)!}}{2^n n!} (-e^{i\phi} \tanh r)^n \ket{2n}.
        \end{split}
    \end{equation}

    The effect of the squeezing operator on the ladder operators is
    \begin{equation}
        \begin{split}
            \mathcal{S}^\dagger(z) \hat{a} \mathcal{S}(z) &= \hat{a} \cosh(r) -\hat{a}^\dagger e^{i \phi} \sinh (r),\\  \mathcal{S}^\dagger(z) \hat{a}^\dagger \mathcal{S}(z) &= \hat{a}^\dagger
        \cosh(r) -\hat{a} e^{-i \phi} \sinh(r),
        \end{split}
    \end{equation}
    and the transformation on the quadrature operators is given by
    \begin{equation}\label{squeezing operator effect}
        \mathcal{S}^\dagger(z) \hat{x}_{\phi} \mathcal{S}(z) = e^{-r}\hat{x}_{\phi}, ~~~ \mathcal{S}^\dagger(z) \hat{p}_{\phi} \mathcal{S}(z) = e^{r}\hat{p}_{\phi}
    \end{equation}
    Using the above transformation, one can compute the mean photon number of a Squeezed state as
    \begin{equation} \label{mu of photon of squeezed}
        \begin{split}
            \bra{z}\hat{n}\ket{z} &= \bra{0}\mathcal{S}^\dagger(z)\hat{a}^\dagger \hat{a} \mathcal{S}(z)\ket{0}\\
            & =\bra{0} \left(\hat{a}^\dagger \cosh(r) - \hat{a} e^{-i\phi}\sinh(r)\right) 
            \left(\hat{a}\cosh(r) - \hat{a}^\dagger e^{i\phi}\sinh(r)\right)\ket{0}\\
            & = \sinh{r}^2,
        \end{split}
    \end{equation}
    and the inner product between two Squeezed states can be calculated as
    \begin{equation}
    \label{squeezed state inner product}
        \braket{z_1,z_2} = \frac{1}{\sqrt{\cosh{r_1}\cosh{r_2} - e^{i(\phi_1 - \phi_2)}\sinh{r_1}\sinh{r_2}}}.
    \end{equation}

\subsection{Displaced Squeezed States}

Combining the action of the Squeezing (\ref{squeezing operator}) and the Displacement operators (\ref{displacement operator}), a Displaced Squeezed State (DSS) which is denoted by $\ket{\alpha, z}$  can be obtained by squeezing a Vacuum state and then displacing it as
\begin{equation}\label{displaced squeezed}
\ket{\alpha, z} = \mathcal{D}(\alpha)\mathcal{S}(z)\ket{0},
\end{equation}
and by using equations \eqref{displacement operator}, \eqref{CS}, \eqref{squeezing operator}, and \eqref{squeezed state representation}, the DSS can be expressed in terms of the Fock basis \cite{book}
\begin{equation}
    \label{DSN}
    \ket{\alpha, z} = e^{-\frac{1}{2}|\alpha|^2-\frac{1}{2}{\alpha^*}^2 e^{i\phi}\tanh{(r)}} \sum_{n=0}^\infty\frac{\left[\frac{1}{2}e^{i\phi}\tanh(r)\right]^{n/2}}{\sqrt{n!\cosh(r)}} H_n\left[ \frac{\alpha\cosh(r)+\alpha^*e^{i\phi}\sinh(r)}{\sqrt{e^{i\phi}\sinh(2r)}} \right]\ket{n},
\end{equation}
since the transformation on the ladder operators is given by
\begin{equation}\label{ladder operator transformation from displaced squeezed state}
    \begin{split}
        \mathcal{D}(\alpha)\mathcal{S}(z) \hat{a} \mathcal{D}^\dagger(\alpha)\mathcal{S}^\dagger(z) & = (\hat{a} - \alpha)\cosh{r} + (\hat{a}^\dagger - \alpha^*)e^{i\phi}\sinh{r}, \\
        \mathcal{D}(\alpha)\mathcal{S}(z) \hat{a}^\dagger \mathcal{D}^\dagger(\alpha)\mathcal{S}^\dagger(z) & = (\hat{a}^\dagger - \alpha^*)\cosh{r} + (\hat{a} - \alpha) e^{-i\phi}\sinh{r}.
    \end{split}
\end{equation}
Similarly, the mean photon number of the DSS can be obtained by a unitary transformation as in equations \eqref{mean photon number coherent} and \eqref{mu of photon of squeezed}
\begin{equation}
    \label{mean photon DSS}
    \bra{\alpha, z}\hat{n}\ket{\alpha, z} = \sinh^2{r} + \vert\alpha\vert^2.
\end{equation}

The inner product between two DSSs is \cite{inner_ds}
\begin{equation}\label{inner displaced squeezed}
    \bra{\alpha_1, z_1}\alpha_2, z_2\rangle = \beta_{21}^{-\frac{1}{2}} \exp\left\{\frac{\gamma_{21} \gamma_{12}^*}{2\beta_{21}} +  \frac{(\alpha_2 \alpha_1^* - \alpha_2^*\alpha_1)}{2} \right\},
\end{equation}
where $\beta_{21} = \cosh{r_2}\cosh{r_1}  - e^{i(\phi_2 - \phi_1)}\sinh{r_2}\sinh{r_1}$ and $\gamma_{21} = (\alpha_2 - \alpha_1)\cosh{r_2} + e^{i\phi_2}(\alpha_2^* - \alpha_1^*)\sinh{r_2}$. It's obvious that when $\alpha=0$, the right-hand side will become identical to equation \eqref{squeezed state inner product} and when $z = 0$ the inner product will  be similar to equation \eqref{cs_inner}. The squeezing part does play an important role in discriminating two quantum states since the CS are not orthogonal from their definition in section \ref{cs}.\\

\subsection{Non-Gaussian States}

From a computational point of view, a generalized DSS can be fully characterized by its Gaussian formalism with the Quadrature operators ordered in a vectorial form and a Covariance matrix that describes the correlations between different modes in a quantum system. This means that a DSS can be efficiently simulated by a classical computer. In Fact, any quadratic Hamiltonian can be decomposed into Squeezing, Displacement, Rotation, and Beamsplitter operators.\\

On the other hand, the Fock state or the Number state is not a Gaussian state. Moreover, higher-order Hamiltonians are required for a universal model of computation. This means that there should be an operator that contains higher-order terms. For instance, the Cubic Phase operator is denoted by $\mathcal{V}(\Gamma) = \exp\left\{i \frac{\Gamma}{3 \hbar} \hat{x}^3\right\}$ where $\Gamma \in \mathbb{R}$. It transforms the lowering operator by the following formula $\mathcal{V}^\dagger(\Gamma) \hat{a} \mathcal{V}(\Gamma) = \hat{a} + i\frac{\Gamma}{2\sqrt{2/\hbar}} (\hat{a} +\hat{a}^\dagger)^2
$. Utilizing such an operator along with the other linear operators will allow us to decompose higher-order Hamiltonians and simulate them efficiently on a photonic quantum computer \cite{Lloyd_1999}.\\
\begin{figure}[!ht]
\centering
\subfigure[Vacuum State]{\label{fig:vac}{\includegraphics[width=0.4\textwidth]{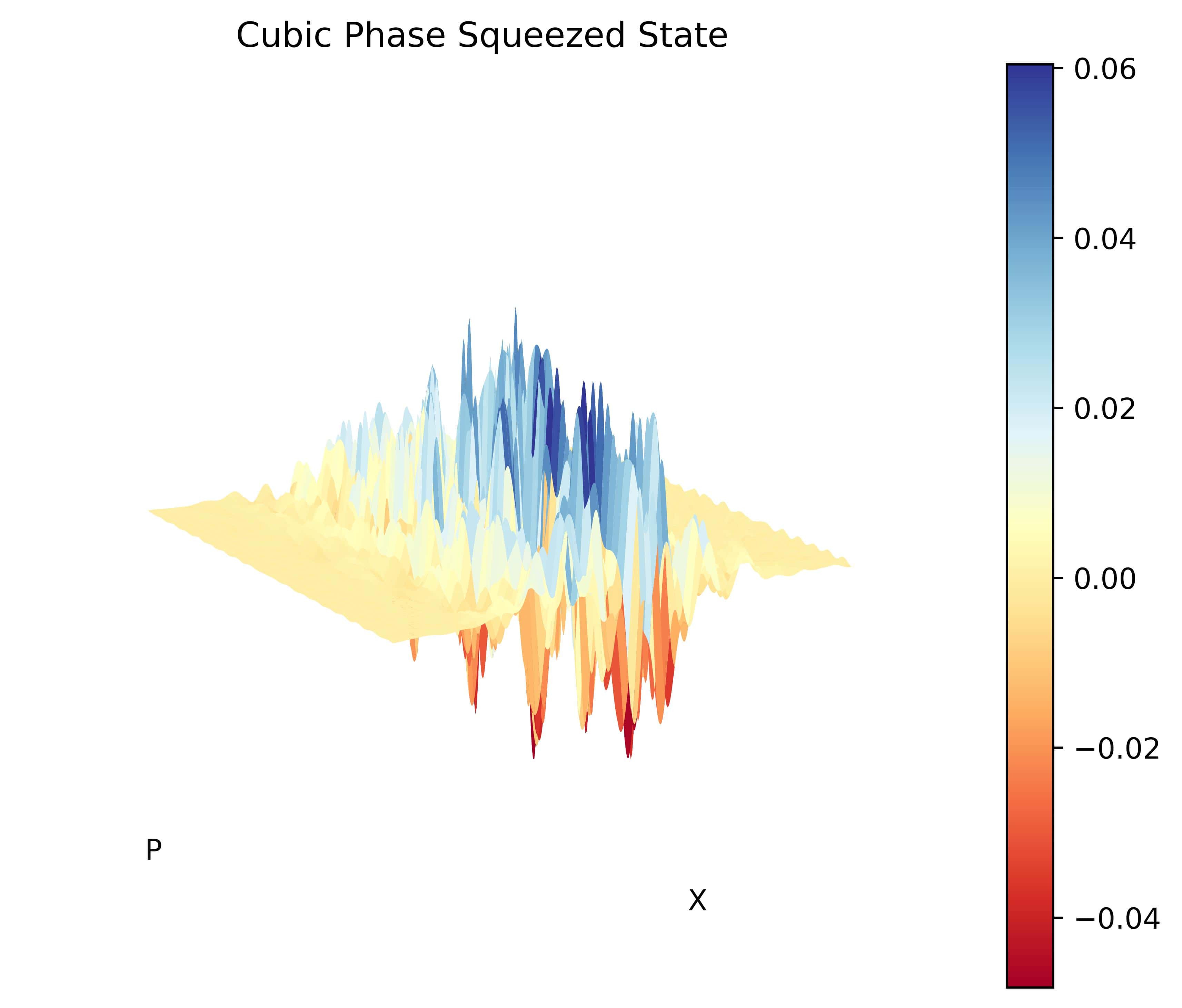}}}\hfill
\subfigure[Fock State of $\ket{1}$]{\label{fig:fock_1}{\includegraphics[width=0.4\textwidth]{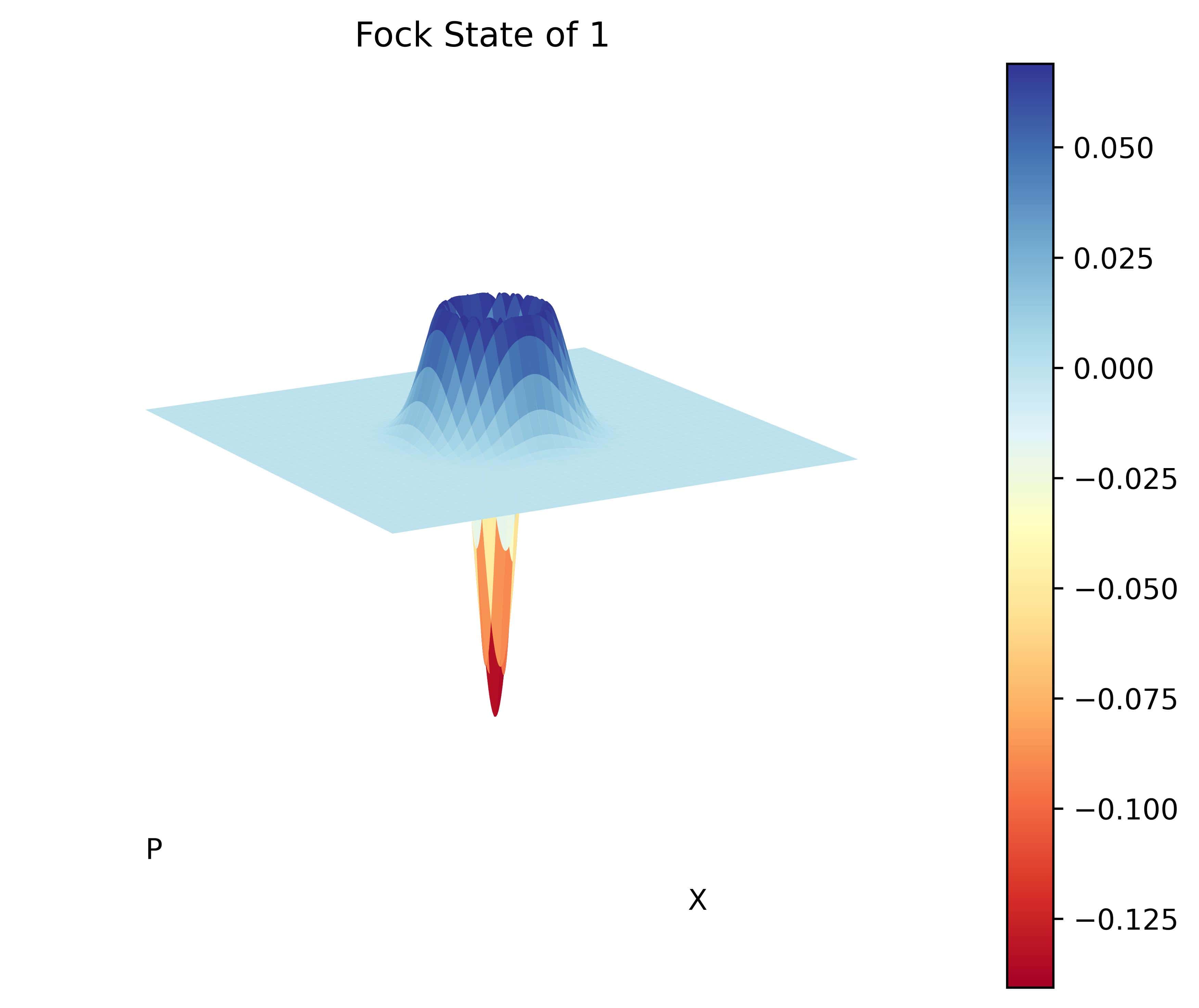}}}
\label{fig:fock}
\vfill
\subfigure[DSS]{\label{fig:dss}{\includegraphics[width=0.4\textwidth]{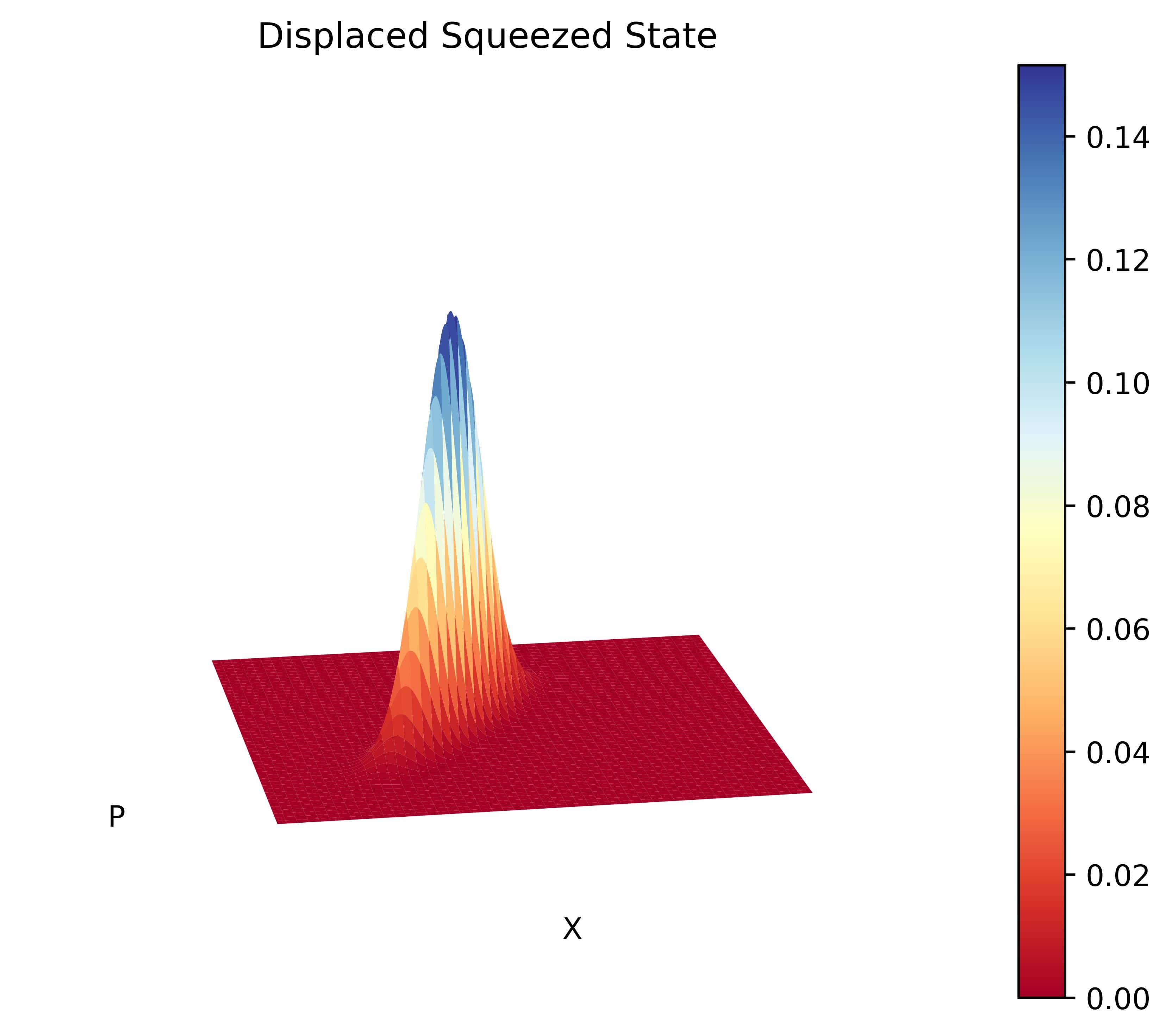}}}
\hfill
\subfigure[A non-gaussian state denoted by $\mathcal{V}(\Gamma)\mathcal{S}(z)\ket{0}$]{\label{fig:cubic}{\includegraphics[width=0.4\textwidth]{paper/assets/cubic_wig-min.jpg}}}
\caption{The Wigner function distributions for different Fock states.}
\label{fig:subfigures}
\end{figure}
To visually express the quantumness arising from non-gaussian operators and to also study different Fock states, the Wigner Function or the \gls{sim}-probability distribution is a systematic method to describe them in terms of the phase-space formalism. Furthermore, the Wigner function is real-valued and encompasses the whole quadrature field operators. It can be calculated using a pure or a mixed state. The Wigner function is defined as
\begin{equation}\label{wigner}
    W(x,p) = \frac{1}{\pi \hbar} \int_{-\infty}^{\infty}
\psi^*(x+y)\psi(x-y)
e^{2i p y / \hbar} dy
\end{equation}
From figure \ref{fig:subfigures} and equation \eqref{wigner}, we can detect a non-gaussian state by looking at the negative part of the \gls{sim}-probability distribution. As expected, Squeezed, Displaced, Displaced Squeezed, and Vacuum states have a positive real-valued outcome. However, any Fock state that has a non-gaussian operation like Cubic phase or Number states, dose have a negative real-valued outcome from the Wigner function.

\subsection{Typical Measurements}
    
\subsubsection{Homodyne and Heterodyne Measurements}
Homodyne and heterodyne measurements are standard detection techniques in optical setups, both of which preserve the Gaussian nature of the state after measurement. Homodyne measurement involves measuring the operator
\begin{equation}
    \hat{x}_{\phi} = \cos (\phi) \hat{x} + \sin (\phi) \hat{p}
\end{equation}
with the resulting probability density given by
\begin{equation}
    p(x_{\phi}) = \langle x_{\phi} | \rho | x_{\phi} \rangle,
\end{equation}
where $x_{\phi}$ represents the eigenvalues of $\hat{x}_{\phi}$. Conversely, in heterodyne measurement, the probability density is given by
\begin{equation}
    p(x_{\phi}) = \frac{1}{\pi} \operatorname{Tr} \left [
        \rho | \alpha \rangle \langle \alpha |
    \right ].
\end{equation}
In optical setups, heterodyne measurement is performed by mixing the state $\rho$ with a vacuum state $\ket{0}$ and then subtracting the detected intensities of the two outputs. The mixing is performed using a 50:50 beamsplitter.

\subsubsection{Particle Number Measurement} \label{app:particle_number}
Particle number measurement, also known as photon detection, is a non-Gaussian projective measurement carried out using number-resolving detectors. The probability of detecting particle numbers $\mathbf{n} = (n_1, n_2, \ldots, n_d)$ is given by
\begin{equation}
    p(\mathbf{n}) = \operatorname{Tr} \left [ \rho | \mathbf{n} \rangle \langle \mathbf{n} | \right ].
\end{equation}
The outcomes are non-negative integer values representing the detected photon numbers.

\section{Devices}
In order to provide initial functionalities, \gls{sim} ships with some idealized versions of quantum devices. In this appendix section we distill some of their implementations.

\subsection{Ideal Fiber}

The most important component in optics simulation is a fiber. Fiber represents a way of connecting different devices in an experiment. The effect of the fiber on the quantum state is in the phase shift, which occurs due to the time difference between input and output of the states. Ideal fiber component is totally defined by its length $l$ and the refractive index of its core $n$. We assume the refractive index of the fiber core $n=1.45$\cite{wang2008investigation}.In addition to that, the ideal fiber is lossless.

With the length and refractive index given, one can compute time $\tau$ each envelope will spend in the fiber. First the speed of light in the fiber can be computed via $c'=\tfrac{c}{n}$ and the time can then be computed with $\tau = \tfrac{l}{c'}$. The time $\tau$ then dictates the phase shift of the quantum state \cite{pegg1997quantum}:

\begin{equation}
    \hat U_F = \exp(i\omega \tau \hat n),
\end{equation}

where the $\omega$ is the central frequency of the photon transversing the fiber \cite{makarov2022theory} and $\hat n = \hat a^\dagger a$ is the number operator. According to this theory we then implement the ideal fiber model.
\begin{codeblock}[htbp]
\centering
\begin{minted}{Python}
class IdealFiber(GenericFiber):
    # ...
    @log_action
    @schedule_next_event
    def des(self, time, *args, **kwargs):
        signals = kwargs.get('signals', {})
        if "length" in signals:
            self.length = float(signals["length"].contents) 
        elif "input" in signals:
            n = 1.45 # refractive index
            t = self.length / (C / n) # tau
            env = kwargs["signals"]["input"].contents # Reference to the Fock space
            op = FockOperation(FockOperationType.PhaseShift, phi=env.frequency * t) 
            env.apply_operation(op) # Application of Phase shift
            return [("output", GenericQuantumSignal().set_contents(env), time + t)]
\end{minted}
\caption{Code for the \py{IdealFiber} class, implementing a phase shift and time delay based on input signals.}
    \label{code:ideal_fiber}
\end{codeblock}

\printbibliography


\end{document}